\documentclass{ltxguide}

\usepackage{color,graphicx,shortvrb}
\usepackage{amsmath,amsthm}
\usepackage{amscd}

\makeindex 

\chardef\bslash=`\\ 

\theoremstyle{definition}

\theoremstyle{remark}

\newcommand{\eval}[2][\right]{\relax
  \ifx#1\right\relax \left.\fi#2#1\rvert}

\DeclareGraphicsExtensions{.ps}

\let\package\textsf

\newlength{\gxlen}
\MakeShortVerb{\|}

\begin{document}

\title{Copernican Cosmography in the First Mexican Physics Treatise}

\author{   Armando  Barra\~n\'on Cedillo
\footnote{ Dept. of Basic Sciences, Universidad Aut\'onoma Metropolitana. Unidad Azcapotzalco.
Av. San Pablo 180, Col. Reynosa-Tamaulipas, Mexico City, 02200. email: bca@correo.azc.uam.mx } } 

\def\rightmark{Nucleaci\'on en la Resistencia Magn\'etica Colosal}
\def\leftmark{A. Barra\~n\'on et al.}

\date{$23^{rd}$, Nov, 2004.}

\maketitle

\abstract
  Copernican Cosmography  was used to discover the path linking  Cebu islands and Acapulco via 
 Kuro-Shivo Stream, as registered in $Physica$ $Speculatio$, first  physics book written in 
New Spain. Teaching and practice of Copernican theory were an outcome of both Spanish
 expansion into the Pacific Ocean and native elite instruction devised by Augustinian 
friars. Copernican cosmography was explained in $Physica$ $Speculatio$'s first 
edition in addition to the reference given by Alonso de la Veracruz about the use of 
Copernican theory to detect Pacific Islands in $Physica$ $Speculatio$'s  Fourth edition. 
Also, Veracroce standpoint about the differences between physics and mathematics is 
discussed in terms of its relation to the rejection of the Copernican Theory as a Natural Theory.

\section{ Introduction}

   In his $Physica$ $Speculatio$, Alonso a Veracroce considered 
solar influence as the cause of several natural events in 
New Spain, namely the abundant and continuous rainfalls at the
 recently conquered Tenochtitlan as well as the eclipses observed in
 restricted regions of the World. Veracroce also explains heliocentric
 theory and the Tychonian system in this production (B. Navarro, 1992:13-24). 
Also, Alonso mentions in the fourth edition of his $Physica$ $Speculatio$,
 the use of the Copernican cosmography in the detection of new islands 
located at the Pacific Ocean. As stated by Alonso de la Veracruz, friar
 Andrea Urdaneta used Copernican cosmography to discover the route
 linking Cebu islands and Acapulco, establishing this way the path 
followed by the so called Manila Nao (Barra\~n\'on, 2004). Therefore, 
we may say nowadays that heliocentric theory was thought and used 
in New Spain at the XVIth century.

 The article is organized in the following way. The second section deals with the
 difussion of Copernican theory in Europe and explains the role played by
 Copernican cosmography in the integration of the Old and New Worlds, connected 
by the new commercial route crossing the Atlantic and Pacific Oceans. 
The third section describes the foundation of the Higher Studies at Tiripet\'{\i}o, 
place where Viceroy Mendoza signed the Treaty that arranged the 
transpacific expedition led by Villalobos. This very section also deals 
with the teaching of Astronomy and Exact Sciences at Tac\'ambaro as part 
of the Augustinian project devoted to native elite education. The fourth 
section analyzes differences between physics and mathematics, as stated
 in the $Physica$ $Speculatio$, which are similar to those argued years later 
in order to declare heliocentric theory as erroneous. Finally, the fifht
 section establishes several conclusions.
\section{ Copernican cosmography and the Unity of the World }

   As Veracroce explains in his $Physica$ $Speculatio$, there is only one world, 
embodying the newly discovered Spanish territories (Alonso de la Veracruz, 
1573:223). Veracroce describes several Spanish settlements in America, 
including their geographic location and makes mention of the mastership of 
the Augustinian friars who were selected as ship pilots for the detection of 
the Pacific Ocean Islands. As explained by Veracroce in the fourth edition
 of his $Physica$ $Speculatio$, the Spanish king had chosen Urdaneta to be
 ship pilot in the exploration of the Pacific Ocean. In order to comply with
 this royal command, Urdaneta used $Alphonsine$ and Copernican Tables to
 delineate the path linking Cebu Islands and Acapulco. As long as these
 islands belonged to Portugal, as signed in Tordesillas Treaty, their 
cosmographic location as well as the discovery of this new route were 
considered of utmost strategic importance for the Spanish crown 
(Alonso de la Veracruz, 1573:223).

   It is easy to understand that Copernican Cosmography was taught in 
New Spain in the XVIth century as long as Copernicus distributed a short version 
of his master work, $De$ $revolutionibus$, ever since the beginnings of that 
century. This $Commentariolus$, arrived to Cracow by 1514, as a theory book
 affirming that the Sun is at rest while the Earth moves around the Sun
 (Noel Swerdlow,1973: 423-512). Besides, it is well known that $Alphonsine$
 $Tables$ were part of the Copernican personal library that included also the 
$Tabulae$ $directionem$ and the $In$ $iudiciis$ $astrorum$ by ar-Rijal (Swerdlow
 y Neugebauer: 3-11). Therefore, when Veracroce wrote his $Physica$ $Speculatio$,
 the Copernican theory was well knwon in Europe and perhaps Veracroce knew 
about it while he was studying and teaching at Salamanca University or in his 
epistolar exchange with Urdaneta.

   The detection of the Manila Nao $tornaviaje$ (return route )demanded the acquaintance with the
 weather and geographical profile of the Pacific Ocean, in order to apply the same
 formula used by the Spaniards to find the return route in the Atlantic Ocean, following
 a northern route. This was the contribution of the Augustinian monk Andrea Urdaneta,
 who benefited from the experience gained from previous California explorations and 
succeeded in discovering the $tornaviaje$ of the Spanish route to the Philippines. 
   Friar Andrea Urdaneta kept a copy of the $De$ $revolutionibus$ and held epistolar 
exchange with Alonso de la Veracruz. Urdaneta received Augustinian orders in 1553 
once he declined the invitation to be general of the Navy destined to discover the 
Western Islands. By 1564 Urdaneta was ship pilot of Legaspi expedition to Cebu
 Islands, departing from Barra de Navidad. When Urdaneta came back from Cebu,
 he found the return route to New Spain, reaching Acapulco in 1565. During this travesy, 
Urdaneta profited from Kuro-Shivo Stream, heading northeast until he reached latitude
 39 N though later Urdaneta turned to latitude 32 N, just to make sure that latitude was
 under control. This way Urdaneta succeeded, after five failures by others, in the detection
 of the return route that provided vast richnesses to Spain ever since the new found lands
 were extremely rich in gold, wax and other luxurious merchandises (Lothar Knauth, 1972).

   As long as the king of Spain had signed Tordesillas Treaty leaving Philippines in hands
 of the Portuguese, he tried to control these islands taking advantage of the fact that
 Philippines were not inhabited by Portugueses at that time. With this purpose, in 1542 the
 Spanish crown organized a military expedition to these islands, baptising them as Philippines
 Islands. Urdaneta had lived nine years in Moluccas Islands, that were recovered from the
 Portugueses in 1535, as demanded by the Spaniards inhabiting them since  Portuguese
 crown had paid 35000 ducados for these islands in 1529 as part of  the Zaragoza Treaty.

   Notwithstanding in 1536 Portugueses stripped Urdaneta of his records and maps relative
 to the journeys of Loaysa and Saavedra, friar Andrea was able to provide an oral version
 of his travel through the Pacific Ocean while he stayed at Valladolid in 1537. After that
 unfortunate event, conquistador Pedro de Alvarado asked Urdaneta to help him in some
 expeditions to New Spain's Pacific coast that Alvarado had in mind. With this purpose,
 Urdaneta embarked in a ship to New Spain in 1538, though Alvarado's death delayed his
 participation in this kind of navigations until he joined Pensacola exploration in 1539. The
 main motivation for the Spanish king to arrange Legaspi's expedition, selecting Urdaneta
 as ship pilot due to his mastering of the Pacific Ocean navigation, was the enforcement of
  Spaniard dominion over the Philippines. This proved to be a convenient strategy when
 in the XVIIth century Portugal forced Spain to accept the border lines in Uruguay, arguing
 that in order to comply with Tordesillas Treaty, Philippines and Moluccas should 
be restored to the Portuguese crown.

   Once the return route linking Cebu and Acapulco was discovered by Urdaneta, an
 intense galleon traffic was established between Philippines and New Spain. Manila
 Nao sailed from Acapulco at a date between the end of Autumn and the beggining 
of Spring. Galleons headed to Cebu once they left Guam, though later they did it to
 Manila Bay. This was a strategic consequence of the use of the 
Copernican cosmography in the delineation of this new commercial route that
 transformed the Spanish king into a Universal Monarch.

   As long as Philippine Nao transported works of art used
 for sumptuary Arts, Mexico city inhabitants celebrated the arrival of the Manila Nao
 ringing the cathedral bells. Galleons transported oriental merchandises such as 
porcelain, regarded at that time as white gold, in volumes simmilar to those carried
 by dutch cargos, which turned New Spain into a commercial emporium (Curiel, 1992).

   The benefits of this Manila Nao's return route impeled the exploration of the Alta
 California by 1570, searching for a safe harbor for the Chinese Nao. Viceroy 
archbishop Moya y Contreras arranged the search for the Armenian islands, probably
 the Hawaiian Archipelago, with the intention to use them as a intermidiate port for the
 Philippine Fleet.

\section{Augustinian  teaching of cosmography in Michoac\'an}

   Fray Alonso de la Veracruz instituted the Colegio Mayor at Tiripet\'{\i}o in 1540, when
 missions departed from Tiripet\'{\i}o and Tac\'ambaro expanding Augustinian territories
 into the Tierra Caliente at the state of Michoac\'an. At the Colegio of Tiripet\'{\i}o, 
students were so well trained that material culture improved in such a way that it was
 possible to build new convents and cities in the surrounding areas though their demand 
deserted the Colegio Mayor. Philosophy and theology were taught for the first time in 
America at Tiripet\'{\i}o, place where Viceroy Mendoza signed a Treaty to explore the Pacific Ocean.
 Besides, in Tac\'ambaro, Alonso de la Veracruz created an astronomical observatory
 and taught therein astronomy and exact sciences (Mauricio Beuchot,16-17). Grijalva
 wrote that the library brought to existence by Veracroce included of maps, celestial
 and terrestrial globes, astrolabes, horologes, $ballestillas$ used to measure angles, 
planispheres and every instrument employed by Liberal Arts.

  This library was formed with the huge collection brought by Alonso from Spain in 
1573 (Juan de Grijalva, fols. 153v y 154r). In this Atlantic journey, Alonso was 
accompanied by nineteen friars to New Spain, as part of his appointment as Visitador
 of the Augustine Congregation in New Spain, Peru and Philippines. In this position he
 was consulted by friar Domingo de Salazar, bishop of Manila, to solve the controversies
 between clerics and regulars (E.J. Burrus, 1968-1976,:63-103).

  By all these contributions to the organization of the scientific studies, Veracroce was
 considered eclesiastic master in New Spain (Sergio M\'endez, 1952:36). As stated by
 Sergio M\'endez Arceo, Augustinian Congregation was leader in the establishment of 
Major Studies in New Spain, excelling other eclessiastical orders that arrived earlier, namely
 dominicans and franciscans. Colegio of Tiripit\'{\i}o received donations from Spaniards and
 natives, founding five haciendas which linked to other plantations formed a passageway 
connecting the Augustinian province. As reported elsewhere, Valladolid's Augustinian
 convent operated as a bank whose influence extended over Zacatula, Zinap\'ecuaro, 
Guanajuato, Tiripet\'{\i}o, Jacona and Ixtl\'an (AHMC, Negocios Diversos, leg. 1 1555-1159).

      Don Antonio Huitzim\'engari, son of Michoacan's king, studied and taught at the Colegio 
Mayor de Tiripet\'{\i}o, as part of the Augustinian project devoted to the education of native elite.
 Huitzim\'engari, who was formerly educated at the franciscan convent of Tzintzuntzan, came 
to Tiripet\'{\i}o in 1540 when Veracroce was principal of the Colegio Mayor, where Huitzim\'engari 
studied sciences, philosophy, theology and languages.

  In 1543, Colegio Mayor moved to Tac\'ambaro while the students missioned into the
 Tierra Caliente twice or thrice a year. And in 1540, Vasco de Quiroga founded the Colegio de 
San Nicol\'as, to educate clerics proficient in native languages. When Alonso de la Veracruz
 governed  once again this province, from 1548 up to 1551, the considerable amount of
 graduates from the Higher Studies helped to institute convents at Guayangareo, Cuitzeo,
 Yuririhap\'undaro, Huango and Charo.
 Therefore, the Augustinian Congregation had already organized the study of cosmography in
 New Spain when Jesuits first arrived to New Spain and were comissioned to administer the
 Colegio de San Nicol\'as in 1574. The arrival of four Jesuits in New Spain had been frustrated  
by illness in 1553 (Juan Moreno, 1965:80). Jesuits received a donation in 1576-77, given by
 the Ayuntamiento of Guayangareo-Valladolid, establishing therein a Colegio. This Colegio
 instructed Grammar to only four locals in 1580-1581 and was supported by the
 generosity of the franciscan an augustinian congregations.

\section{Heliocentrism as a Natural Theory}

 Heliocentrism adheres to Aristotleian cosmology inasmuch as Copernicus 
assumes a natural motion of Earth and planets,  considers that Sun remains still, includes
 celestial spheres and regards a stationary higher sphere at the universe's border. 
Copernicus is unique in terms of his abbility to reach an agreement among Neoplatonics,
 Phytagoreans and Cabbalists. In spite of the reputation earned for his mathematicial
 resources, when Copernicus places the Sun in the center of the planetary motions 
where the Sun should be best placed  to enlighten the planets, Copernicus recollects 
Hermes Trimegistus who called the Sun a visible God ( N. Cop\'ernico, 1873: 15-17).
  In regards to solar influence, Veracroce follows a natural perspective to deal with the
 eclipses and local time delays between America and Europe. For example, Alonso a 
Veracroce uses the circular shape of the Earth to explain what we call today time zones
 as well as the fact that eclipses are only visible in some regions of the planet. Veracruz
 uses this geometrical scheme to explain local time delays between Mexico City 
and Toledo  (Alonso de la Veracruz, 1573:223).

   Alonso recognizes the failure of Aristotle's $Meteorology$ Treatise, whose Second Book
 predicts that excesive sunlight should avoid the presence of water, snow and abundant
 vegetation in the torrid zone. Instead of, Alonso reports copious and 
continous rainfalls as well as snow production nearby Mexico City (Alonso de la Veracruz, 1573:281).

    In agreement with this natural perspective, Alonso denies any astrological influence on
 terrestrial events, but accepts solar lightening as a natural cause, rejecting all grounds
 to the $Treatise$ $Disputationes$ $adversus$ $astrologiam$ $divinatricem$ written by Pico 
de la Mir\'andola (Alonso de la Veracruz, 1573:281).  

  Alonso explains heliocentric theory in Chapter  XVII  of the $Treatise$  $on$ $Sphere$ written
 by Campano, telling that some people consider that Earth is always orbiting and fixed 
Stars stand still (Bernab\'e Navarro, 1992:13-24), namely a theoretical postulate of the
 $Commentariolus$ (Alfred Romer,1999: 157-183).  Veracroce's explanation of 
 heliocentrism is supplemented by his expounding of the $Tychonian$ system in $Physica$ 
$Speculatio$'s Chapter 52 (Bernab\'e Navarro, 1992:13-24).

    Copernicus used mathematical arguments to ground a System of the World, in spite
 of the hierarchy of disciplines accepted at that time, where physics occupied a higher place
 than mathematics. That's why Giovanni Maria Tolossani criticized  $De$ $Revolutionibus$, 
since Tolossani disregarded  Copernicus as expert in physics and Holy Scripture
 (E. Garin, 1976:288). 

   In a simmilar fashion, Cardinal Bellarmino considered physical hypothesis as mere 
computational premises, establishing an important antecedent to the rejection of 
heliocentrism as a theory of Nature (A. Koestler, 1986:I 125). These were the premises 
assumed to classify heliocentrism as erroneous in 1616, besides the authenticity 
declaration of the Vulgata Latina at the Tridentine Council.

  Alonso Guti\'errez considers substantial differences between physics and mathematics, 
as stated in his $Physica$ $Speculatio$. They are different in regards to the object of study 
as well as to the level of abstraction. This way, Veracroce adheres to the standpoint held
 by Roman College, that sustained the incapability of mathematics to provide a thorough
 image of the World.

\section{Conclusions.}

  As declared by Alonso de la Veracruz, Copernican cosmography was
 used to discover the path linking  Cebu islands and Acapulco via the
 Kuro-Shivo Stream. Also, Copernican cosmography was explained in the
 $Physica$ $Speculatio$, first  physics book written in New Spain. Teaching 
and practice of Copernican theory were a byproduct of the overseas Spanish
 expansion as well as of the native elite instruction devised by the Augustinian 
friars. Notwithstanding $Physica$ $Speculatio$ is devoted to Aristotelian and Ptolomeian 
cosmography, Copernican theory is already explained in $Physica$ $Speculatio$ first 
edition and the practical use of Copernican theory to detect the $tornaviaje$ is mentioned 
in its Fourth edition. Nevertheless, Veracroce accepts that physics and mathematics 
employ different abstraction levels and are dedicated to distinct objects of study. This way, 
Veracroce implicitly supports the standpoint of the Roman College, that finally declared 
Copernican theory as erroneous inasmuch as a mathematics was considered unable at
 that ime to provide a natural theory.

\section{Acknowledgments}

 Author acknowledges hospitality from the Mexican 
National Library at UNAM (Fondo Reservado) and financial support from 
UAM -Azcapotzalco.

\end{document}